\documentclass[aps,showpacs,prc,nofootinbib,floatfix,]{revtex4}

\usepackage{graphicx}

\usepackage{latexsym}
\usepackage{hhline}
\usepackage{amssymb}
\usepackage{wasysym}
\usepackage{epsfig}

\sloppypar

\newcommand{\be}{\begin{equation}}
\newcommand{\ee}{\end{equation}}
\newcommand{\bea}{\begin{eqnarray}}
\newcommand{\eea}{\end{eqnarray}}
\newcommand{\ba}{\begin{array}}
\newcommand{\ea}{\end{array}}
\newcommand{\bi}{\begin{itemize}}
\newcommand{\ei}{\end{itemize}}
\newcommand{\mi}{\mbox i}

\newcommand{\refe}[1]{(\ref{#1})}

\newcommand{\mck}{{\mathcal K}}
\newcommand{\mcl}{{\mathcal L}}

\newcommand{\mct}{{\mathcal T}}

\newcommand{\ra}{\rightarrow}

\renewcommand{\slash}{/ \!\!\!\!\,}

\newcommand{\foh}{\frac{1}{2}}

\newcommand{\fth}{\frac{3}{2}}
\newcommand{\ffh}{\frac{5}{2}}

\newcommand{\umat}{1 \! \! 1}

\newcommand{\JPR}[3]{Phys. Rev. {\bf #1}, #2 (#3)}
\newcommand{\JPS}[3]{{Phys. Scr.} {\bf #1}, #2 (#3)}
\newcommand{\JPRL}[3]{{Phys. Rev. Lett.} {\bf #1}, #2 (#3)}
\newcommand{\JPL}[3]{{Phys. Lett.} {\bf #1}, #2 (#3)}
\newcommand{\JPRC}[3]{Phys. Rev. C {\bf #1}, #2 (#3)}
\newcommand{\JPRD}[3]{Phys. Rev. D {\bf #1}, #2 (#3)}
\newcommand{\JNP}[3]{Nucl. Phys. {\bf #1}, #2 (#3)}

\newcommand{\JZP}[3]{{Z. Phys.} {\bf #1}, #2 (#3)}

\newcommand{\Jpin}[3]{{$\pi N$-Newsletter} {\bf #1}, #2 (#3)}

\newcommand{\ibid}[3]{{ibid}, {\bf #1}, #2 (#3)}

\newcommand{\JPRep}[3]{{Phys. Rept.} {\bf #1}, #2 (#3)}

\begin{document}

\title{ Spin-$\ffh$ resonance  effects in a coupled channel nucleon resonance analysis
\footnote{Supported by DFG and GSI Darmstadt}}

\author{V. Shklyar}
\email{shklyar@theo.physik.uni-giessen.de}
\author{G. Penner}
\author{U. Mosel}
\affiliation{Institut f\"ur Theoretische Physik, Universit\"at Giessen, D-35392
Giessen, Germany}

%\date{\today}

\begin{abstract}
Spin-$\ffh$ resonance contributions in pion-nucleon scattering 
are investigated within the coupled channel $K$-matrix 
approach for c.m. energies up to $\sqrt s=$ 2 GeV.
 All previously studied $\pi N$, $2\pi N$,
$\eta N$, $K \Lambda$, $K \Sigma$, and $\omega N$ final states are included.
We find a significant improvement of  $\chi^2$ in almost all channels by inclusion of 
the  spin-$\ffh$ states. 
The obtained coupling  parameters are discussed.
\end{abstract}


\maketitle

\section{Introduction}
The extraction of  baryon-resonance properties is one of the important 
tasks of modern hadron physics. Great efforts have been made in the past to 
obtain this information from the analysis of pion- and photon-induced  reaction data. 
The precise knowledge of 
these properties is an important step towards understanding the hadron structure and 
finally the strong interactions. However, until now there are states  whose 
properties are not settled or whose existence is rather 
controversial. At the same time even some resonances with four-star status have 
their parameters in a wide range as extracted from different analyses \cite{pdg}.  
This complicates the comparison of the extracted parameters with theoretical model 
predictions. On the other hand,  some quark models (see \cite{Capstick02} 
and Refs. therein) predict  
that the  baryon resonance spectrum may be richer then discovered  so far. 
This is a problem of 'missing' nucleon resonances. One assumes that these states
are weakly coupled to  pion channels and 
are consequently not clearly seen in  $\pi N$ and $2\pi N$ reactions from 
which experimental
data are most often used for baryon-resonance analyses. To solve these problems
additional final states must be taken into account. Ideally a comprehensive
analysis should include all open channels and take all experimental  reaction data
into account.

A large number of models were suggested to obtain resonance properties  by using the 
experimental data from  hadronic reactions only (\cite{arndt95,manley92,vrana}; 
see  also \cite{pdg} for references).
All these analyses apply different unitarization methods for the $T$-matrix  and  
deal with various data sets. In  \cite{arndt95} the 
authors use  only the $\pi N$ data where inelastic reactions are 
approximated  by a ``generic'' $\pi \Delta$ channel. The analysis of 
Manley and Saleski \cite{manley92} uses 
both $\pi N$ and $2\pi N$ production data  whereas the most recent studies of 
Vrana et al. \cite{vrana} include $\pi N \to\eta N$ cross sections data in addition to the 
last two ones.  To incorporate other possible finale states a unitary coupled channel 
model (Giessen model) has been developed which includes  $\gamma N$, $\pi N$, $2\pi N$, 
$\eta N$, $K \Lambda$  final states and deals with all available experimental
data on pion- and  photon-induced reactions \cite{feusti98,feusti99}. Most recent
 extensions
of this model include $K \Sigma$ and $\omega N$ final states \cite{Penner02,pm2} as well, 
which allows for the simultaneous analysis of all hadronic and photoproduction data 
up to $\sqrt s =$ 2 GeV. 
A shortcoming of this study is the missing of higher spin resonances with spin 
$J> \fth$. A successful description of data for all final states has nevertheless been achieved 
in \cite{Penner02} and  at  first sight  there is not much evidence that higher-spin
resonances can give significant contributions to the final states studied. However, for a 
comprehensive analysis of baryon spectra the contributions from higher spin resonances 
must be included as well. 
It is  
known as well, that \mbox{spin-$\ffh$}  resonances have large electromagnetic couplings \cite{pdg} and have
to be included into photon-induced reaction analyses. 
To improve the situation we extend our model by including resonance states  
with spin $\ffh$. In the present paper we report our first results on the pion-induced
 reactions studied; the simultaneous analysis  of  pion- and photon-induced  
reactions will be presented in a subsequent paper.

As a first step we keep all ingredients of our last study taking into account the
$\pi N$, $2\pi N$, $\eta N$, $K \Lambda$, $K \Sigma$, and $\omega N$ final states and 
include in addition the spin-$\ffh$ resonances that were absent in 
\cite{feusti98,feusti99,Penner02,pm2}. We do not look for  additional open 
channels but instead
check the evidence for additional resonance states which have a one or two 
stars status rated by \cite{pdg}. We start in Sec. \ref{model} with 
a description of the formalism  concentrating  mainly on 
the spin-$\ffh$ couplings; the complete discussion of our model including 
all other couplings can be found in \cite{Penner02,pm2,phd}. Sec. \ref{details} is devoted
to details of calculations.
In  Secs. \ref{reslt} and \ref{ExtrPar} we discuss the results of our calculations in comparison with 
the previous studies \cite{Penner02}. 
We present the extracted resonance masses and partial decay widths and finish with a summary.

\section{The Giessen model}
\label{model}
We solve the Bethe-Salpeter coupled-channel equation in the $K$-matrix approximation  
to extract scattering amplitudes for the final states under consideration. In order to decouple
the equations we perform  partial-wave decomposition of the $T$ matrix into good 
total spin $J$, isospin $I$, and parity $P=(-1)^{J\pm\frac{1}{2}}$. Then the partial-wave amplitudes 
can be expressed in terms of an interaction potential $\mck$ via the matrix equation
\bea
\mct^{I,J\pm} = 
\left[ \frac{\mck^{I,J\pm} }{1 - \mi \mck^{I,J\pm}}\right], \;
\label{bsematinv}
\eea
where each element of the matrices $\mct^{I,J\pm}_{fi}$ and $\mck^{I,J\pm}_{fi}$ 
corresponds to a given initial and final state ($i,f =$ $\pi N$, $2\pi N$, $\eta N$, 
$K \Lambda$, $K \Sigma$, $\omega N$ ). The interaction potential is approximated 
by tree-level Feynman diagrams which in turn are obtained from effective 
Lagrangians \cite{Penner02,phd}. The $\mct$-matrix \refe{bsematinv}  fulfils  
unitarity  as long as the $\mck$ matrix is hermitian.

The Lagrangian for the spin-$\ffh$ resonance decay to a final baryon $B$ and a (pseudo)scalar  
meson 
$\varphi$ is chosen in the form
\bea
\mcl^{\ffh}_{\varphi B R} = \frac{{\rm g}_{\varphi B R}}{m_\pi^2}\bar u_R^{\mu \nu}
\Theta_{\nu\lambda}(a) \Gamma_S   u_B  \partial_\mu \partial^\lambda 
\varphi + h.c.
\label{Lagran52}
\eea
with the  matrix $\Gamma_S=\umat$ if resonance and final meson have identical parity  and  
$\Gamma_s=\mi\gamma_5$  otherwise. The free spin-$\ffh$ Rarita-Schwinger symmetric field  
$u_R^{\mu \nu}$ obeys the Dirac equation and satisfies the conditions 
$\gamma_\mu u_R^{\mu \nu}=\partial_\mu u_R^{\mu \nu} ={\rm g}_{\mu\nu} u_R^{\mu \nu} =0$ 
\cite{rarita}.

The off-shell projector $\Theta_{\mu\nu}(a)$
is defined by 
\bea
\Theta_{\mu\nu}(a) = {\rm g}_{\mu\nu} - a\gamma_\mu\gamma_\nu,
\label{off}
\eea
where $a$ is the off-shell parameter. 
In general the interaction Lagrangian \refe{Lagran52}
can have two off-shell projectors matched with both vector indices of the resonance field tensor. 
However, as we will see later, a good description of the experimental
 data can be achieved already with a single 
parameter $a$ keeping 
the second one equal to zero. Thus we have found no necessity for additional parameters  and
to keep our model as simple as possible we use only  one off-shell projector in \refe{Lagran52}. 

The widths of the hadronic resonance decays as extracted from the Lagrangian  
\refe{Lagran52}
are 
\bea
\Gamma_\pm(R_\ffh\to\varphi B)= I\frac{{\rm g^2_{\varphi B R}}}{30\pi m_\pi^4}k_\varphi^5
\frac{E_B\mp m_B}{\sqrt{s}}.
\label{width}
\eea
The upper sign corresponds to the decay of the resonance into a meson with the  
identical parity and
vice versa. $I$ is the isospin factor and $k_\varphi$, $E_B$, and $m_B$ are the meson  momentum, 
energy and mass of the final baryon, respectively. 

The coupling  of the spin-$\ffh$ resonances to the   
$\omega N$ final state is chosen  to be 
\bea
\mcl^{\ffh}_{\omega N} =
\bar u_R^{\mu \lambda}\Gamma_V
\left( \frac{{\rm g}_1}{4m_N^2}\gamma^\xi
+\mi\frac{{\rm g}_2}{8m_N^3}\partial^\xi_{N}
+\mi\frac{{\rm g}_3}{8m_N^3}\partial^\xi_{\omega}\right)
(\partial^{\omega}_\xi{\rm g}_{\mu\nu} - 
\partial^{\omega}_\mu{\rm g}_{\xi\nu})u_N \partial_\lambda^\omega  \omega^\nu + h.c.,
\label{LgrOm}
\eea
where the matrix $\Gamma_V$ is $\umat$ ($\mi\gamma_5$) for positive (negative) resonance parity 
and $\partial^\mu_N$ ($\partial_\mu^\omega$)  denotes the partial derivative of
the nucleon  and the $\omega$-meson  fields, respectively. 
The above Lagrangian is constructed in 
the same manner as the one for  spin-$\fth$ in \cite{Penner02}. Similar couplings were also used to 
describe 
electromagnetic processes \cite{Wolf,David,Hand,Titov}.   
Since the different parts of \refe{LgrOm} contribute at different kinematical conditions we 
keep all three couplings as  free parameters  and vary them during the fit. 
As stressed in  \cite{Penner02} the couplings in \refe{LgrOm}  can not be reliably derived 
from hadronic data only; photoproduction data are required as an additional constraint to
fix the constants.
Therefore,  in the present study 
 we do not try to fix all $\omega N$ couplings
but look for the total $\omega N$ flux contribution for each resonance state. 
To save calculation time and for the sake of simplicity,  we also 
do not introduce
any off-shell parameters at the $\omega$-meson couplings \refe{LgrOm}; indeed we found no 
strong contribution from spin-$\ffh$ waves to the $\omega$-production channel.
The couplings \refe{LgrOm} lead to the helicity-decay amplitudes 
\bea
A^{\omega N}_{\fth} &=&
\frac{\sqrt{E_N\pm m_N}}{\sqrt{5m_N}}
\frac{k_\omega}{4m_N^2}
\left (
-{\rm g}_1(m_N\mp m_R)
+{\rm g}_2\frac{(m_R E_N-m_N^2)}{2m_N}
+{\rm g}_3\frac{m_\omega^2}{2m_N^2} \right ),  \nonumber\\
A^{\omega N}_{\foh} &=&
\frac{\sqrt{E_N\pm m_N}}{\sqrt{10m_N}}
\frac{k_\omega}{4m_N^2}
\left ( 
{\rm g}_1 (m_N\pm (m_R-2E_N)) 
+{\rm g}_2\frac{(m_R E_N -m_N^2)}{2m_N}
+{\rm g}_3\frac{m_\omega^2}{2m_N^2} \right ),  \nonumber\\
A^{\omega N}_{0}&=&\frac{\sqrt{(E_N\pm m_N)}}{\sqrt{5m_N}}
\frac{k_\omega m_\omega}{4m_N^2} 
\left ( {\rm g}_1
\pm {\rm g}_2\frac{E_N}{2m_N}
\pm {\rm g}_3\frac{(m_R-E_N)}{2m_N} \right ),
\label{helic}
\eea
with upper (lower) signs corresponding to positive (negative) resonance parity. The lower 
indices stand for the helicity $\lambda$ of the final $\omega N$ state
$\lambda=\lambda_V -\lambda_N$ where 
we use  an
abbreviation as follows: $\lambda=$ $0: 0+\foh$, ~$\foh:1-\foh$,
~ $\fth:1+\foh$. 
The resonance $\omega N$ 
decay width $\Gamma^{\omega N}$ is written as the sum over the three helicity amplitudes given above:
\bea
\Gamma^{\omega N}= \frac{2}{(2J+1)}\frac{k_\omega m_N}{2\pi m_R}
\sum_{\lambda=0}^{3/2} |A^{\omega N}_\lambda|^2,
\label{helicW}
\eea
where  $J=\ffh$ for the spin-$\ffh$ resonance decay. 

For practical calculations we adopt the spin-$\ffh$ projector in the form 
\bea
P_{\ffh}^{\mu\nu,\rho \sigma}(q)&=&\frac{1}{2}(
 T^{\mu\rho}T^{\nu \sigma}
+T^{\mu \sigma}T^{\nu \rho} )
-\frac{1}{5}2T^{\mu\nu}T^{\rho \sigma}\nonumber\\
&+&\frac{1}{10}(
T^{\mu\lambda} \gamma_\lambda\gamma_\delta T^{\delta \rho}T^{\nu \sigma}
+T^{\nu\lambda} \gamma_\lambda\gamma_\delta T^{\delta \sigma}T^{\mu \rho}
+T^{\mu\lambda} \gamma_\lambda\gamma_\delta T^{\delta \sigma}T^{\nu \rho}
+T^{\nu\lambda} \gamma_\lambda\gamma_\delta T^{\delta \rho}T^{\mu \sigma}),
\label{ConvPr}
\eea
with
\bea
T^{\mu\nu}&=&-{\rm g}^{\mu\nu}+\frac{q^\mu q^\nu}{m_R^2},
\label{ConvPr2}
\eea
which has also  been  used in an analysis of  $K\Lambda$
photproduction \cite{David}. 
As is well known the description of particles with spin  
$J>\foh$ leads to a number of different propagators which have 
non-zero  off-shell lower-spin  components. To control these components the off-shell 
projectors \refe{off} are usually introduced. There were attempts to fix the off-shell
parameters and remove the spin-$\foh$ contribution in the 
case of spin-$\fth$ particles \cite{nath}.
However, it has been shown \cite{ben89} that these contributions cannot be suppressed for any 
value of $a$. To overcome this problem Pascalutsa suggested
gauge invariance as an 
additional constraint to fix the interaction Lagrangians for higher spins and remove
the lower-spin components \cite{pascatim}. 
Constructing the spin-$\fth$ interaction for a Rarita-Schwinger field $u^\mu_{\fth}$
by only  allowing couplings to the gauge-invariant field 
tensor  $U^{\mu\nu}_{\fth}=\partial^\mu u^\nu_{\fth}-\partial^\nu u^\mu_{\fth}$  Pascalutsa  
derived an interaction  which (for example) for the $\pi N \Delta$ coupling is
\bea
\mcl_{\pi N\Delta}=  f_\pi \bar u_N\gamma_5 \gamma_\mu \widetilde{U}^{\mu\nu}
\partial_\nu \varphi + h.c.
\label{Psc}
\eea
where $\widetilde{U}^{\mu\nu}$ is the tensor dual to $U^{\mu\nu}$: 
$\widetilde{U}^{\mu\nu} = \varepsilon^{\mu\nu\lambda\rho}U_{\lambda\rho}$ and 
$\varepsilon^{\mu\nu\lambda\rho}$ is the Levi-Civita tensor. 
The same arguments can also be applied to spin-$\ffh$ particles. In this case the
amplitude of meson-baryon scattering can be obtained from the  conventional amplitude 
by the  replacement 
\bea
\Gamma_{\mu\nu}(p',k')\frac{P_{\ffh}^{\mu\nu,\rho \sigma}(q)}{\slash q-m_R}
\Gamma_{\rho\sigma}(p,k)
\to
\Gamma_{\mu\nu}(p',k')\frac{{\mathcal P}_{\ffh}^{\mu\nu,\rho \sigma}(q)}{\slash q-m_R}
\Gamma_{\rho\sigma}(p,k)\frac{q^4}{m_R^4},
\label{PscCoupl}
\eea
where  $\Gamma_{\rho\sigma}(p,k)$ are vertex functions that follow from 
\refe{Lagran52} and \refe{LgrOm} by applying  Feynman rules
 and the projector ${\mathcal P}_{\ffh}^{\mu\nu,\rho \sigma}(q)$
is obtained from (\ref{ConvPr}, \ref{ConvPr2}) by the replacement 
$q^\mu q^\nu/m_R^2 \to q^\mu q^\nu/q^2$.
This procedure is similar to that which has been used in the spin-$\fth$ case 
\cite{pascatim}.  
As can be  clearly seen, the conventional and the Pascalutsa descriptions  give the same 
results for on-shell particles. 
It has been shown  for the spin-$\fth$ case  \cite{pasca01} ,  that both prescriptions are 
equivalent in the effective Lagrangian approach as long as additional contact interactions
are taken into account when the Pascalutsa couplings are used.  
The differences between these descriptions  have been discussed in 
\cite{pascatjon,Penner02,lahiff} and here we perform calculations by using both approaches.

In order to take into account the internal structure of mesons and baryons each  vertex is 
dressed by a corresponding formfactor:
\bea
F_p (q^2,m^2) &=& \frac{\Lambda^4}{\Lambda^4 +(q^2-m^2)^2}. 
\label{formfacr} 
\eea
Here  $q$ is the four momentum of the intermediate particle and $\Lambda$ is a cutoff parameter.
In \cite{Penner02} it  has been shown that the
 formfactor \refe{formfacr} gives  systematically better 
results  as compared to other ones, therefore we do not use any other forms for $F(q^2)$. The 
cutoffs $\Lambda$ in \refe{formfacr} are treated as  free parameters and allowed to be varyed 
during the  calculation. However we demand the same cutoffs in all channels for a 
given  resonance
spin  $J$ : $\Lambda^{J}_{\pi N}=\Lambda^{J}_{\pi\pi N}=\Lambda^{J}_{\eta N}=...$ etc., 
($J=\foh,~ \fth,~ \ffh$). This greatly reduces the number of free parameters; i.e. for all 
spin-$\ffh$ resonances there is  only one cutoff $\Lambda_{\ffh}$ for all decay channels.

To take into account  contributions of the $2\pi N$ channel in our calculations
we use the inelastic partial-wave cross section $\sigma_{2\pi N}^{JI}$ data 
extracted  in \cite{manley84}. To this end the inelastic $2\pi N$ channel is parameterized by 
an effective $\zeta N$ channel where $\zeta$ is an effective isovector meson with mass 
$m_\zeta=2m_\pi$.
Thus the $\zeta N$ is considered as a sum of different
($\pi \Delta$, $\rho N$, etc.) contributions to the total partial-wave $2\pi N$ flux. 
 We allow
only resonance $\zeta N$-couplings since 
each background diagram would introduce
a meaningless coupling parameter.
Despite  this approximation the studies 
\cite{feusti98,feusti99,sauermann,Penner02} have achieved a good description of the total partial 
wave cross sections \cite{manley84} and we proceed in our calculations by using the above 
prescription. For the $R \ra \zeta N$ interaction  the same Lagrangians are used as for
the $R \ra \pi N$ couplings  taking into account the positive parity of the $\zeta$ meson.

\section{Details of calculations}
\label{details}
In our  previous work  \cite{Penner02}  a good description of all hadronic experimental data
was achieved. Hence the best hadronic fit results from 
\cite{Penner02} 
for the conventional ( C-p-$\pi+$) and the Pascalutsa ( P-p-$\pi+$ ) prescriptions 
are used as  starting points in our extended calculations. We apply the same database as 
in \cite{Penner02} with additional elastic $\pi N$ data for the spin-$\ffh$ partial wave 
amplitudes taken from the VPI group analysis \cite{SM00}. For the $2\pi N$ channel we
 use the  spin-$\ffh$ partial  wave cross sections derived in \cite{manley84}. Due to the lack 
of data for higher energies  we confine ourselves to the energy region 
 $m_\pi + m_N \leqslant \sqrt{s} \leqslant 2$ GeV.
\begin{table}
  \begin{center}
    \begin{tabular}
      {l|r|r|r|r|r|r|r}
      \hhline{========}
       Fit & Total $\pi$ & $\chi^2_{\pi \pi}$ & $\chi^2_{\pi 2\pi}$ & 
       $\chi^2_{\pi \eta}$ & $\chi^2_{\pi \Lambda}$ & $\chi^2_{\pi \Sigma}$
       & $\chi^2_{\pi \omega}$ \\ 
       \hline
       C-p-$\pi +$ ($\ffh$) & 2.60 & 2.60 & 7.63 & 1.37 & 2.14 & 1.83 & 1.23 \\
       C-p-$\pi +$ & 2.66 & 3.00 & 6.93 & 1.85 & 2.19 & 1.97 & 1.24 \\
       P-p-$\pi +$ ($\ffh$) & 3.65 & 3.80 &10.06 & 1.75 & 2.54 & 2.93 & 1.83\\
       P-p-$\pi +$ & 3.53 & 3.72 & 9.62 & 2.47 & 2.69 & 2.92 & 2.17 \\
      \hhline{========}
       C-p-$\pi +$ ($\ffh$)$^*$ & 2.46 & 2.70 & 7.11 & 1.37 & 2.14 & 1.83 & 1.23 \\
       P-p-$\pi +$ ($\ffh$)$^*$ & 3.37 & 3.72 & 9.22 & 1.75 & 2.54 & 2.93 & 1.83\\
       \hline
    \end{tabular}
  \end{center}
  \caption{$\chi^2$ of the different fits. The first and third  lines
  are the new best fit results including the spin-$\ffh$ resonances. 
  The second and fourth lines correspond to results from our previous study \cite{Penner02}.
  $^*$: $\chi^2$ is calculated neglecting experimental  spin-$\ffh$ partial wave data in $\pi N$ and
  $2\pi N$ channels. In all cases the $D_{35}$ data have not been taken into account
   (see text). 
    \label{sqr}}
\end{table}

A complete discussion of the fitting procedure can be found in \cite{phd}.
Here we briefly give the main features of the calculations. During  the fit
a $\chi^2$ minimization was performed, where $\chi^2$ is defined as
\bea
\chi^2 = \frac{1}{N} \sum\limits_{n=1}^N 
\left( \frac{x_c^n - x_e^n}{\Delta x_e^n} \right)^2 \; .
\label{hi}
\eea
Here, $N$ is the total number of datapoints, $x_c^n$ and $x_e^n$ are calculated and experimental
values  and  $\Delta x_e^n$ is the experimental uncertainty of $x_e^n$.
To reduce the
number of free parameters  the $\pi N$ and $2\pi N$ subset of 
the spin-$\ffh$ resonance  parameters was varied first 
to find a satisfying description of the $\pi N$ and $2\pi N$ data keeping  all 
resonance and background couplings from C-p-$\pi+$ and P-p-$\pi+$  fixed. 
Since the Pascalutsa coupling is free of any lower-spin off-shell contributions the inclusion
of additional spin-$\ffh$ resonances does not lead to any modification of the  
$\pi N$ and $2\pi N$ lower partial-waves description. 
On the other hand the conventional
couplings  give rise to the  lower $\pi N$ and $2\pi N$ partial waves 
and also  modify $\eta N$, $K\Sigma$, $K\Lambda$ and $\omega N$ amplitudes
due to  rescattering effects even for zero couplings in these channels. 
Therefore, the simultaneous variation of the spin-$\fth$ and -$\ffh$ off-shell parameters 
is also needed. 

After a preliminary fitting of the spin-$\ffh$-resonance masses and $\pi N$ and $2\pi N$ 
parameters  we carry out an overall fit for all non- and resonance couplings  in comparison  
with the available experimental hadronic data.  To minimize the number 
of free  parameters  we vary only the two coupling constants ${\rm g}_1$,  ${\rm g}_2$ in 
the Lagrangian  \refe{LgrOm}.
As we will see later,  in  Sec. \ref{reslt}, a successful description of 
$\omega$-meson data can be achieved by using these two couplings only.  

The resulting $\chi^2$ of our best 
overall hadronic fits are given in Table \ref{sqr} in comparison with our previous results
(last two lines). Here we keep the notation of \cite{Penner02}:  C-p-$\pi +(\ffh)$ 
and  P-p-$\pi +(\ffh)$ are the best new fit results for the conventional and the Pascalutsa 
coupling calculations. The third and fourth lines show the $\chi^2$ from the previous study.
 We find a problem with the description of the $D_{35}$ partial wave so  
the resulting $\chi^2_{\pi \pi}$
turns out to be very large (see the discussion in the next section). 
Hence  $\chi^2_{\pi \pi}$ values given in Table \ref{sqr}   are  calculated by neglecting  the
$\pi N$ datapoints for the $D_{35}$ partial wave.

One sees from Table \ref{sqr} that the inclusion of the spin-$\ffh$ resonances leads to
an improvement
of $\chi^2$ in almost all channels for calculations using the conventional spin-$\ffh$ 
couplings. The only exception is the $2\pi N$ channel  where the increase in 
$\chi^2_{\pi 2 \pi}$ is mainly  related to the description of 
the $D_{15}$ and $F_{15}$ partial waves (see next section). Calculations using 
the Pascalutsa prescription lead to a  $\chi^2$ at almost the same level of quality as for
our previous results \cite{Penner02}. 
The obtained new $\chi^2_{\pi \pi}$ and $\chi^2_{\pi 2\pi}$ are 
calculated using experimental data  from all $\pi N$ and $2\pi N$  partial waves
up to spin-$\ffh$. For a more
reliable comparison with previous results, we have calculated $\chi^2_{\pi \pi}$ and  
$\chi^2_{\pi 2\pi}$  when only the spin-$\foh$ and -$\fth$ partial wave data were taken into
account in \refe{hi}. The obtained values for the conventional and the Pascalutsa 
coupling calculations \mbox{(last two lines in  Table \ref{sqr})}  show that the inclusion 
of  spin-$\ffh$ resonances  improves the description of the experimental data.  

\section{Results and discussion}
\label{reslt}

\begin{figure}
  \begin{center}
    \parbox{16cm}{
      \parbox{75mm}{\includegraphics[width=70mm]{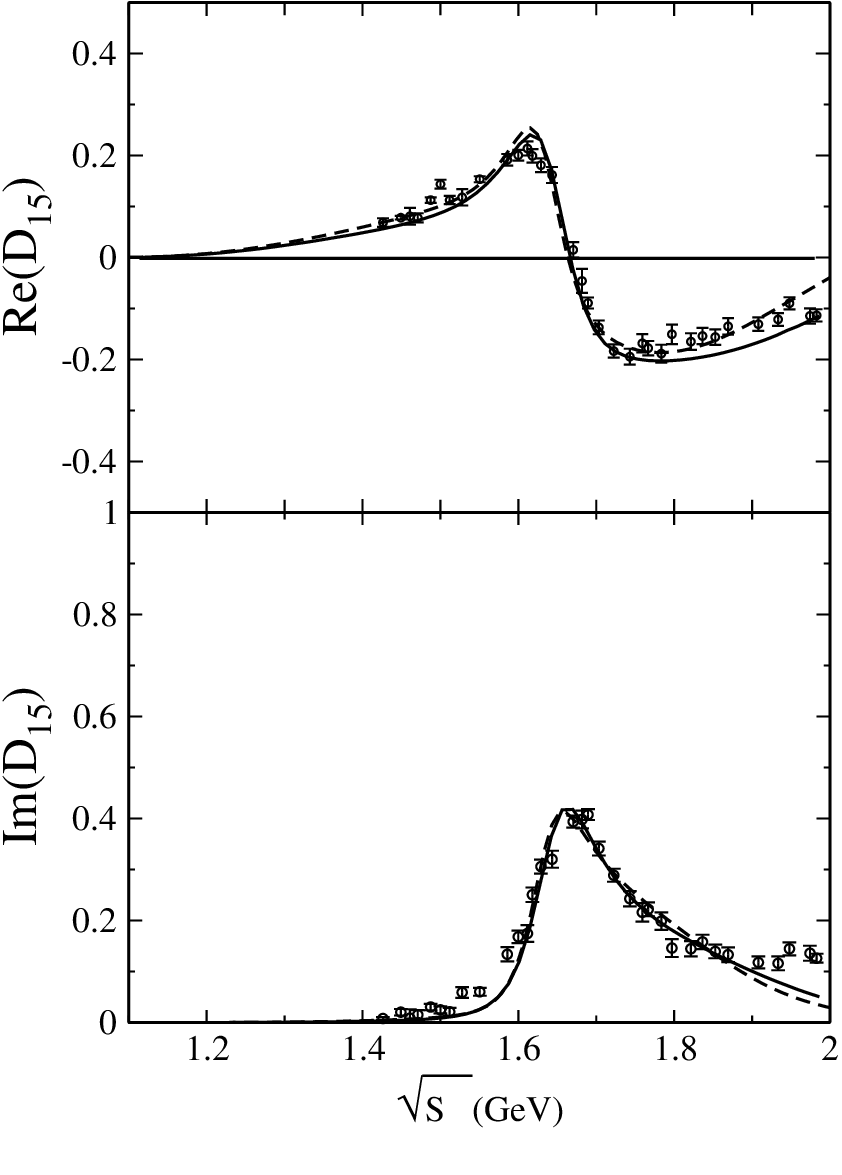}}
      \parbox{75mm}{\includegraphics[width=70mm]{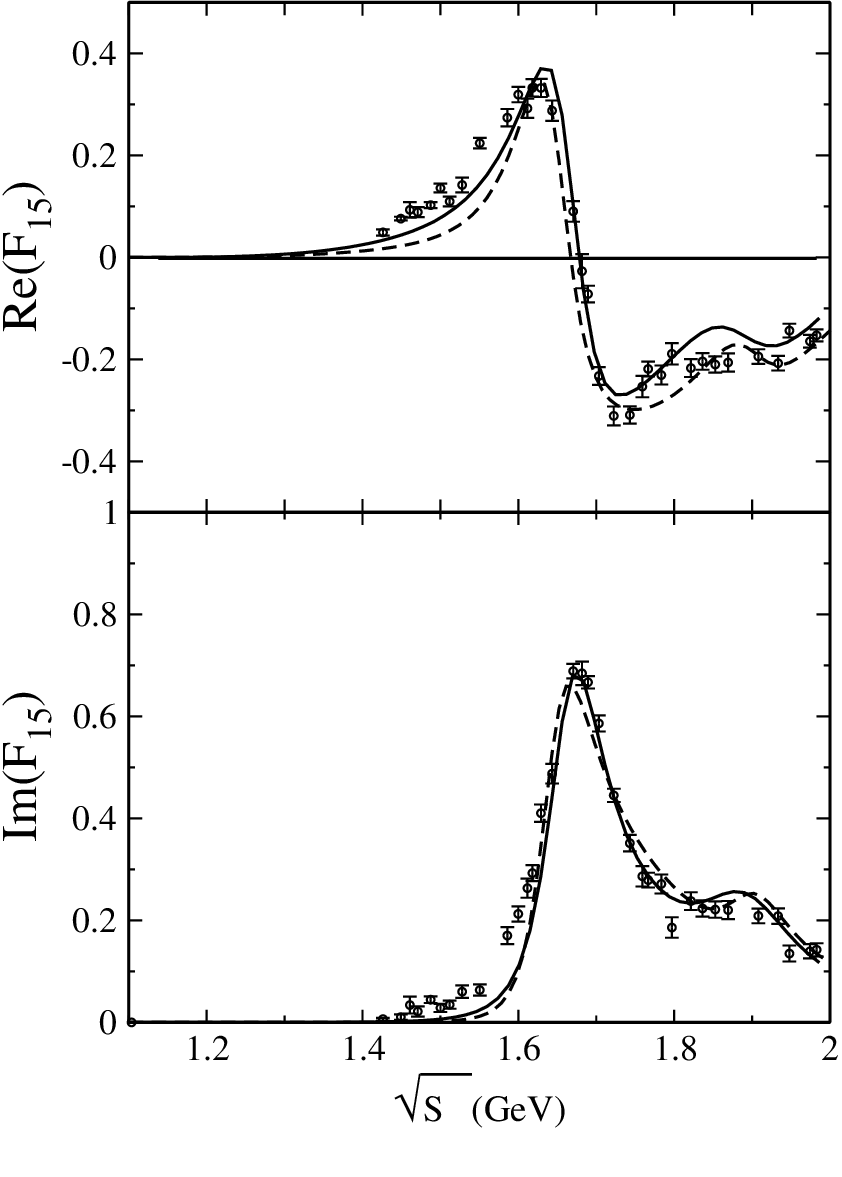}}
       }
       \caption{ 
The elastic $\pi N \to \pi N$ scattering amplitudes for the  spin-$\ffh$ partial waves.
The solid (dashed) line  corresponds to calculation  C-p-$\pi +(\ffh)$ 
(P-p-$\pi +(\ffh)$). The data are taken from \cite{SM00}. 
      \label{d15f15}} 
  \end{center}
\end{figure}
\begin{figure}
  \begin{center}
    \parbox{16cm}{
      \parbox{75mm}{\includegraphics[width=70mm]{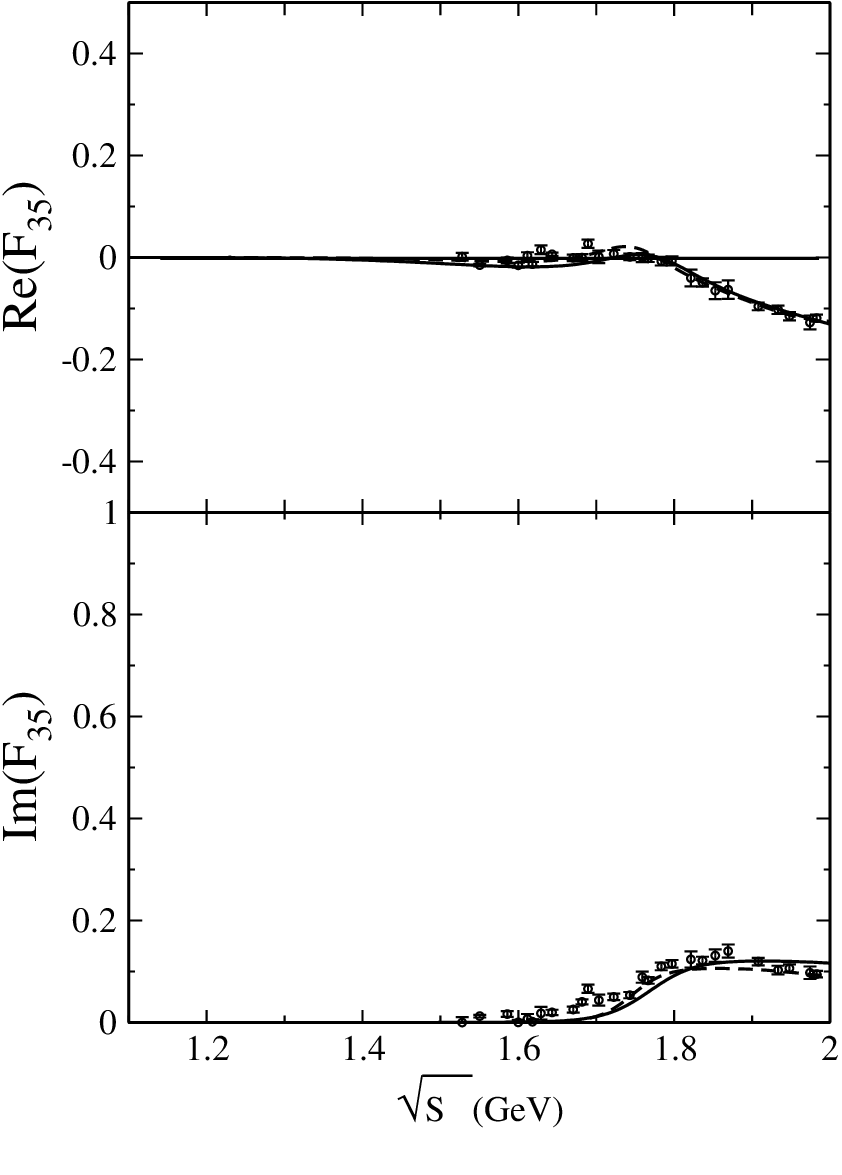}}
      \parbox{75mm}{\includegraphics[width=70mm]{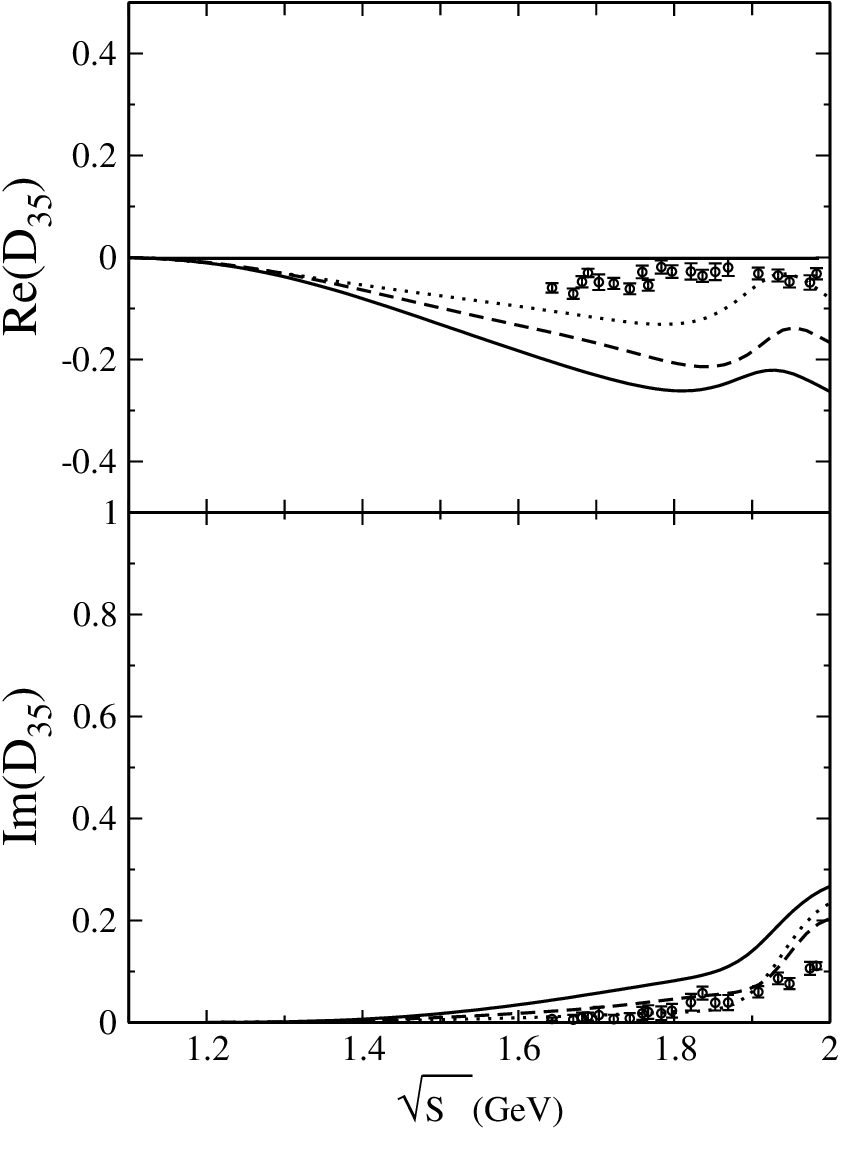}}
       }
       \caption{ 
The elastic $\pi N \to \pi N$ scattering amplitudes for the  spin-$\ffh$ partial waves.
The solid (dashed) line  corresponds to calculation  C-p-$\pi +(\ffh)$ 
(P-p-$\pi +(\ffh)$). The dotted line in the right figure is the result for reduced nucleon
cutoff (see text). The data are taken from \cite{SM00}. 
      \label{d35f35}} 
  \end{center}
\end{figure}
\begin{figure}
  \begin{center}
    \parbox{16cm}{
      \parbox{75mm}{\includegraphics[width=70mm]{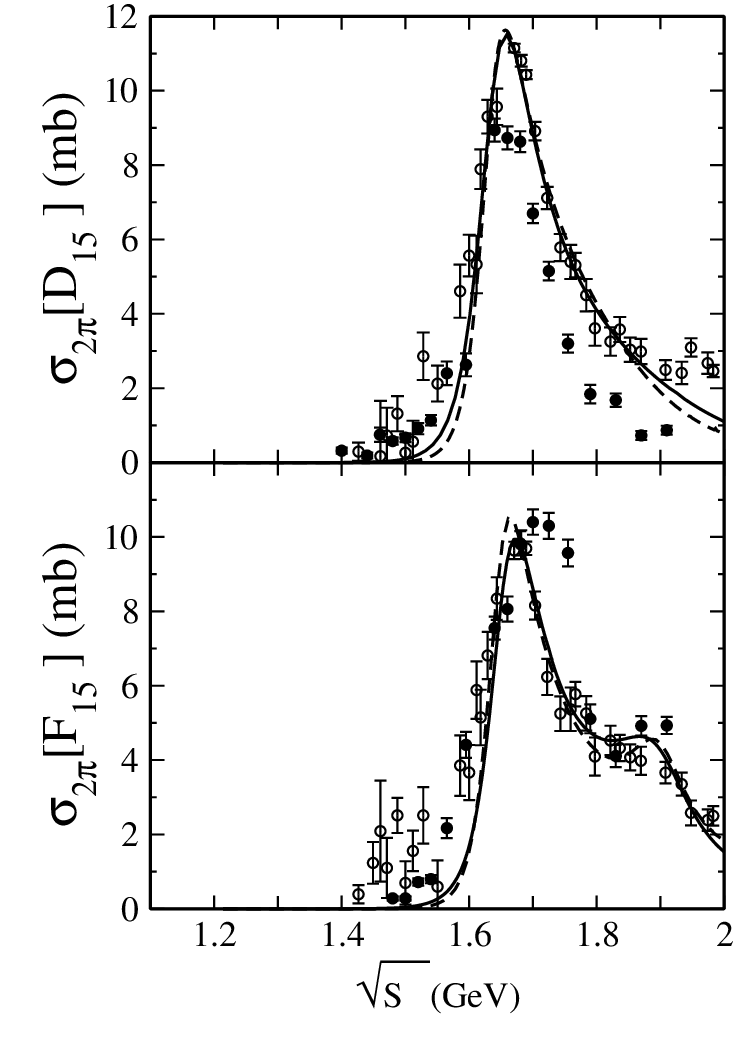}}
      \parbox{75mm}{\includegraphics[width=70mm]{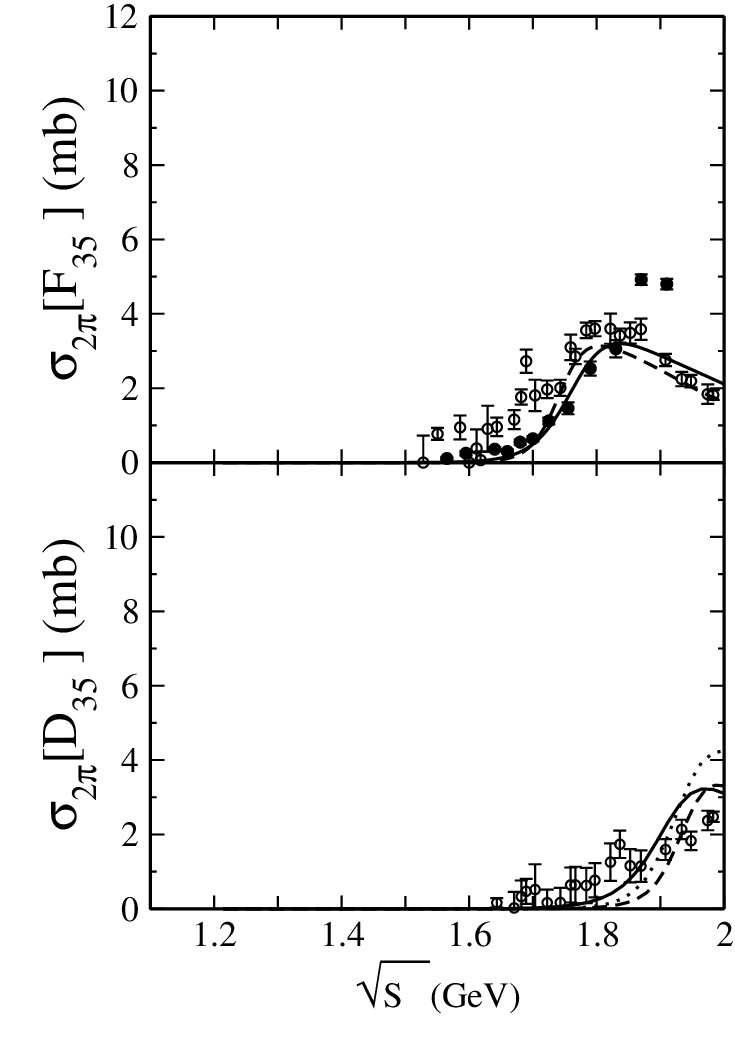}}
       }
       \caption{ 
The inelastic $D_{15}$, $F_{15}$, $F_{35}$, and $D_{35}$ waves. The solid (dashed) line  
corresponds to calculation  C-p-$\pi +(\ffh)$ (P-p-$\pi +(\ffh)$)  for the $2\pi N$ channel.
Open and filled circles represent
the total inelasticity from the VPI group \cite{SM00} and the $2\pi N$ data \cite{manley84}, 
respectively. The calculated inelasticities 
almost coincide with the calculated $2\pi N$ cross sections  and are  not shown here. 
Calculation with a reduced nucleon cutoff is shown by the dotted line.
      \label{fig2}} 
  \end{center}
\end{figure}

In this section we discuss  the results of our calculations for the spin-$\ffh$ waves.
We  stress that the $\pi N$ partial  wave inelasticities are not fitted  
but  obtained as a sum of the individual contributions from all open channels.  
The  $D_{15}(1675)$,
$F_{15}(1680)$,  and  $F_{35}(1905)$ resonances were included in our 
calculations. We have also found evidence for a second  $F_{15}$ state around 1.98 GeV which 
is rated two-star by \cite{pdg}. The results for the elastic $\pi N \to \pi N$ amplitudes and 
$\pi N \ra 2\pi N$ partial-wave cross  sections are shown in Figs. \ref{d15f15} - \ref{fig2}

The $\pi N$ and $2\pi N$ channels are found to be the dominant decay modes for all 
four states.
In the following each spin-$\ffh$ wave is discussed  separately.

{$\bf D_{15}$}. The elastic VPI data show a single resonant peak which corresponds  
to the well established  $D_{15}(1675)$ state. We find a
good description of the elastic amplitude in  both the C-p-$\pi +(\ffh)$ and the
P-p-$\pi +(\ffh)$  calculations. 

The $2\pi N$ data  \cite{manley84} are systematically below the total inelasticity of the VPI 
group \cite{SM00}.
 This can be an indication that apart from  $2\pi N$  there are 
additional contributions from other inelastic channels. 
However, in the analysis of Manley and Saleski 
\cite{manley92} as well as in the most recent 
study of Vrana et al. \cite{vrana}  the total inelasticity in the $D_{15}$ wave is
entirely  explained by the resonance decay to the $\pi \Delta$ channel. We also find no
significant  contributions from the $\eta N$, $K\Lambda$, $K\Sigma$, and $\omega N$ channels
to the total $\pi N$ inelasticity in the present hadronic calculations. 
The calculated $2\pi N$ cross sections are found to be substantially above the data 
from \cite{manley84} in all fits. 
Indeed, the difference between the $2\pi N$  and  inelasticity 
data runs into 2 mb at 1.67 GeV. This flux can be absorbed by neither $\eta N$, $K\Lambda$, 
$K\Sigma$, $\omega N$ channels, see Fig.\,\ref{crstot}. Thus we conclude that either the 
$\pi N$ and $2\pi N$ data are inconsistent with each other or other open channels 
(as 3$\pi N$) must be taken 
into account. To overcome this problem and to describe the $\pi N$ and $2\pi N$ data in 
the $D_{15}$  partial wave the original $2\pi N$ data error bars \cite{manley84} 
 were weighted by a
factor 3. The same procedure was also used by Vrana et al. \cite{vrana} and Cutkosky et al. 
\cite{cutkosky90} to fit the inelastic data.

In both conventional and Pascalutsa coupling calculations the  total inelasticities in the  
$D_{15}$ wave almost coincide with the partial wave cross sections and therefore are not 
shown in Fig.\,\ref{fig2} (left top). A good description of the inelasticity in the $D_{15}$ 
wave was achieved and the extracted resonance parameters are also in agreement with other 
findings (see next section). 

{$\bf F_{15}$}. The  $F_{15}(1680)$ and $F_{15}(2000)$ resonances are identified in 
this partial  wave. 
The inclusion of the second  resonance significantly improves the description of the
$\pi N$ and $2\pi N$ experimental
data in the higher-energy region. Some evidence for this state was  also found in earlier 
works \cite{manley92,hohler}. 
A visible inconsistency  
between the inelastic VPI data and  the 
$2\pi N$ cross section from \cite{manley84} above 1.7 GeV  can  be seen in Fig. \ref{fig2} 
(left bottom). 
The three  data points at 1.7, 1.725, and 1.755 GeV have, therefore, not been included in our 
calculations. 
Finally we achieve a reasonable description for both $\pi N$ and $2\pi N$ data. The 
conventional and Pascalutsa coupling calculations give approximately the same results.  

{$\bf F_{35}$}. A single resonance state $F_{35}(1905)$ was taken into account.
Some other  models find an additional lower-lying resonance with a mass of about 1.75 GeV
\cite{manley92,cutkosky79,hohler,vrana}. 
However, we already find a good description of the elastic 
$\pi N$ amplitudes and the $2\pi N$ cross sections by only including the single $F_{15}(1905)$ 
state. 
The inclusion of a second state with somewhat lower mass leads to a worse
description of the $\pi N$ and $2\pi N$ data due to the strong interference between 
the nearby states. 
The two $2\pi N$ data points at 1.87 and 1.91 GeV,  which are apparently above the total 
inelasticity have  not been  included into calculations.   

The total inelasticity in the $F_{35}$ partial wave almost coincides 
with the calculated
$2\pi N$ cross section and is not shown in Fig.\,\ref{fig2}. Note, that the $2\pi N$ data 
at 1.7 GeV are slightly below the total inelasticity from
\cite{SM00}. This could indicate that other inelastic channels give additional contributions
to this partial wave.

There are also difficulties in the description of the $2\pi N$ low-energy tails of the $D_{15}$ 
and $F_{15}$ partial waves below 1.6 GeV, where the calculated cross sections
are slightly below the $2\pi N$ data. The discrepancy leads to a significant rise in 
$\chi^2_{\pi 2\pi}$ (cf. Table \ref{sqr}). The same  behavior 
has been found in our previous calculations for the $D_{I3}$ partial waves \cite{Penner02}. 
There, it has been suggested that the problem might be caused by the description of the
$2\pi N$ channel in terms of an effective $\zeta N$ state. Indeed,
the findings of \cite{manley92,vrana} show   strong $\pi \Delta$ decay ratios  in 
all three
$D_{15}$,  $F_{15}$, and  $F_{35}$ partial waves. The description of the $2\pi N$ channel in terms 
of the $\rho N$ and $\pi \Delta$  channels may change the situation when
taking into account the
$\rho N$ and $\pi \Delta$ phase spaces and corresponding spectral functions. 
Upcoming calculations will address this question.

{$\bf D_{35}$}.  A single $D_{35}(1930)$ resonance is taken into account.
However, there is no clear resonance structure in  the $\pi N$ data for this partial wave.
The data \cite{SM00} also show a total inelasticity at the 2 mb 
level whereas the $2\pi N$ channel was found to be negligible \cite{manley84}. 
It has been suggested \cite{manley84} that this channel could have an important  
inelastic $3\pi N$ contribution. 
Since the measured $2\pi N$ cross section is zero we have used the inelastic $\pi N$ data with
enlarged error bars instead of  the $2\pi N$ data to pin down the $2\pi N$ $D_{35}$ contributions. 
Even in this case  we have found difficulties in the description of the $D_{35}$ partial wave. 
The $\pi N$ channel turns to be strongly influenced by the $u$-channel nucleon and 
resonance contributions which give significant contributions to the real part of $D_{35}$. 
As can be seen in Fig.\ref{d35f35} the conventional and Pascalutsa coupling calculations 
cannot give even a rough description of the experimental data \cite{SM00}. 
The situation can be improved by either using the reduced nucleon cutoff $\Lambda_N$ 
or by neglecting the nucleon $u$-channel contribution in the interaction kernel. 
The latter approximation has been used  in the coupled-channel approach of Lutz et al. \cite{lutz}. 
To illustrate this point we have carried out an additional fit for the reduced cutoff 
 $\Lambda_N$=0.91 taking 
only the $\pi N$ and $2\pi N$ data into account. The calculated  $\chi^2$ are 
$\chi^2_{\pi \pi}$=3.63 and  $\chi^2_{\pi 2\pi}$=7.87 where the $D_{35}$ data are also taken 
into account in \refe{hi} (note, that all values in Table\,\ref{sqr} are calculated 
by neglecting these datapoints ). The results for the $D_{35}$ partial wave are shown in 
Fig.\,\ref{d35f35} by the dotted line. In all calculations for $D_{35}$ presented in 
Fig.\,\ref{d35f35} the $D_{35}(1930)$ mass was found to be of about 2050 MeV. 
One sees that the calculations with  a reduced nucleon cutoff lead to a better 
description of the $D_{35}$ data giving, however, a worse description of other 
$\pi N$ partial  
wave data. Note, that a reduction of the nucleon cutoff is required for a  successful 
description of the lower-spin photoproduction multipoles \cite{Penner02,pm2}
which also leads to  a worsening in $\chi^2$ for the $\pi N$ channel. 
Here we leave the problem of description of the $D_{35}$ partial wave
for future study when the photoproduction data will be also included 
thus allowing more constrain for the background contributions.

Finally, we  conclude that the main features of the  considered spin-$\ffh$ partial  waves 
except $D_{35}$  are well reproduced. 
From \mbox{  Figs. \ref{d15f15}-\ref{fig2}}
one can see that there is no significant difference between the conventional 
\refe{ConvPr} and the Pascalutsa \refe{PscCoupl} spin-$\ffh$ couplings. The results 
for the lower-lying spin-$\foh$ and -$\fth$ partial waves are only slightly changed from
our previous study \cite{Penner02} and we do not show them here.

\section{Extracted resonance  parameters}
\label{ExtrPar}
\subsection{Spin-$\ffh$ states}
\label{paramtr}

\begin{table}[t]
  \begin{center}
    \begin{tabular}
      {l|c|r|r|r|r|r|r|r}
      \hhline{=========}
      $L_{2I,2S}$ & mass & $\Gamma_{tot}$ &
      $R_{\pi N}$ & $R_{2\pi N}$ & $R_{\eta N}$ &
      $R_{K \Lambda}$ & $R_{K \Sigma}$ & $R_{\omega N}$ \\
      \hhline{=========}
$D_{15}(1675)$& 1665 & 144 & 40.2 & $59.1(-)$ & $ 0.6(-)$ & $ 0.0(+)$ & $-0.04^a$ & --- \\
              & 1662 & 138 & 41.2 & $58.4(+)$ & $ 0.4(-)$ & $ 0.0(-)$ & $ 0.02^a $ & --- \\
      \hhline{=========}
$F_{15}(1680)$& 1674 & 120 & 68.5 & $31.5(-)$ & $ 0.1(+)$ & $ 0.0(+)$ & $ 0.07^a$ & --- \\
              & 1669 & 122 & 65.8 & $34.2(+)$ & $ 0.0(-)$ & $ 0.0(+)$ & $ 0.13^a$ & --- \\
     \hline
$F_{15}(2000)$& 1981 & 361 &  9.0 & $84.0(+)$ & $ 4.3(-)$ & $  0.5^b(-) $ & $ 0.4(-)$ &  2.2 \\
              & 1986 & 488 &  9.5 & $88.2(-)$ & $ 0.3(-)$ & $ 0.1(+)$ & $ 0.2(-)$ &  1.7 \\
      \hhline{=========}
$F_{35}(1905)$& 1859 & 400 &  11.3 & $ 88.7(+)$ & --- & ---  & $ 0.7^b(+)$ & --- \\
              & 1830 & 457 &  10.3 & $ 89.7(-)$ & --- & ---  & $ 0.0(+)$ & --- \\
      \hhline{=========}
    \end{tabular}
  \end{center}
  \caption{
Properties of the  spin-$\ffh$ resonances considered in the
    present calculation.
    The first line corresponds to calculation C-p-$\pi +(\ffh)$ and the second to 
    P-p-$\pi +(\ffh)$.
    Masses and total widths $\Gamma_{tot}$ are given 
    in MeV. The decay ratios $R$ are shown in percent of the total width.
    $^a$: The coupling is presented since the resonance is
    below threshold. 
 $^b$: Decay ratio in
    0.1\permil. 
    \label{52param}} 
\end{table}

\begin{figure}
  \begin{center}
    \parbox{16cm}{
      \parbox{75mm}{\includegraphics[width=69mm]{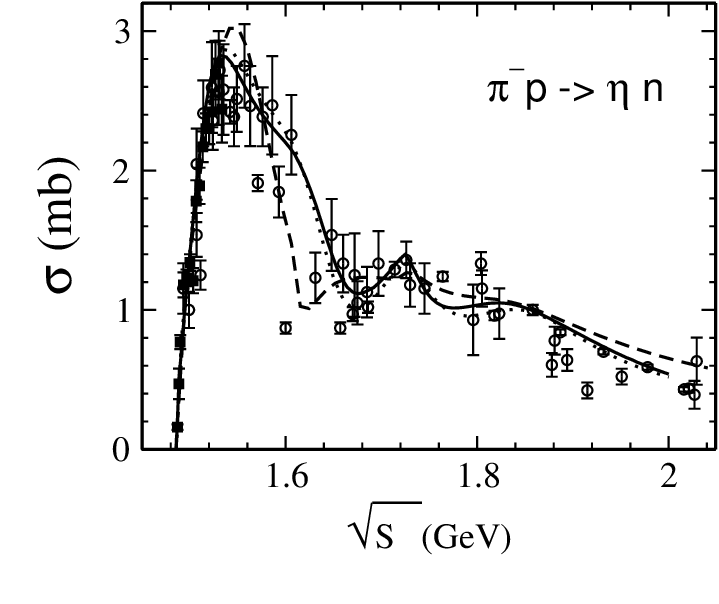}}
      \parbox{75mm}{\includegraphics[width=69mm]{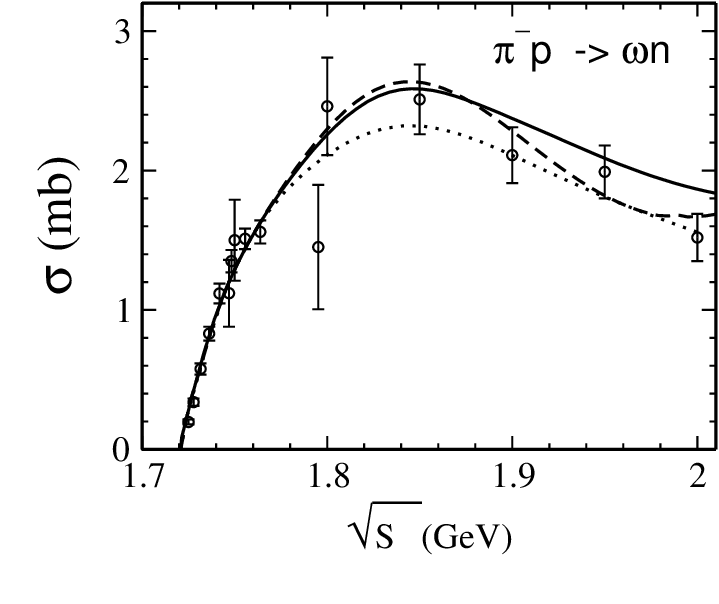}}
       }\vspace*{0.1cm}
    \parbox{16cm}{
      \parbox{75mm}{\includegraphics[width=69mm]{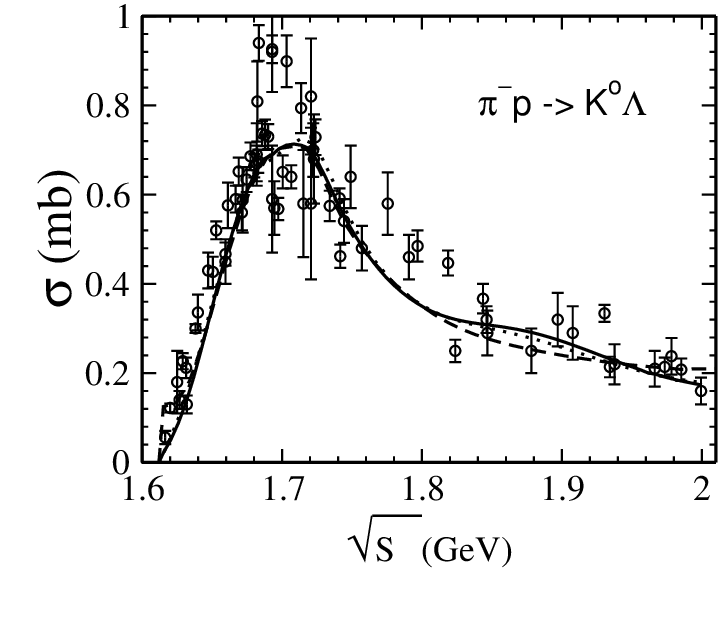}}
      \parbox{75mm}{\includegraphics[width=69mm]{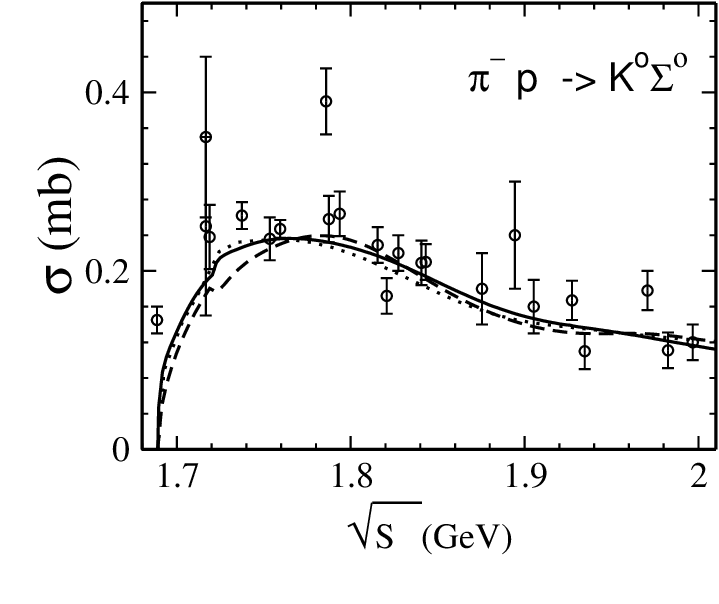}}
       }\vspace*{0.1cm}
    \parbox{16cm}{
      \parbox{75mm}{\includegraphics[width=69mm]{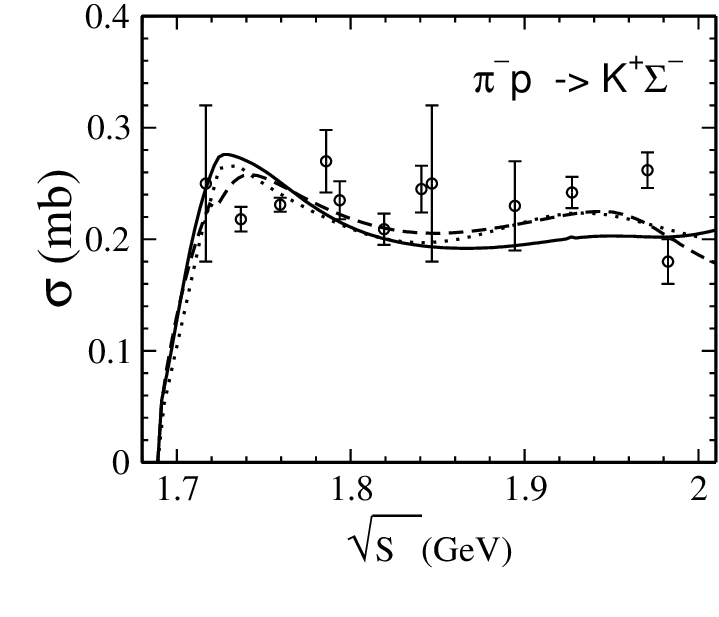}}
      \parbox{75mm}{\includegraphics[width=69mm]{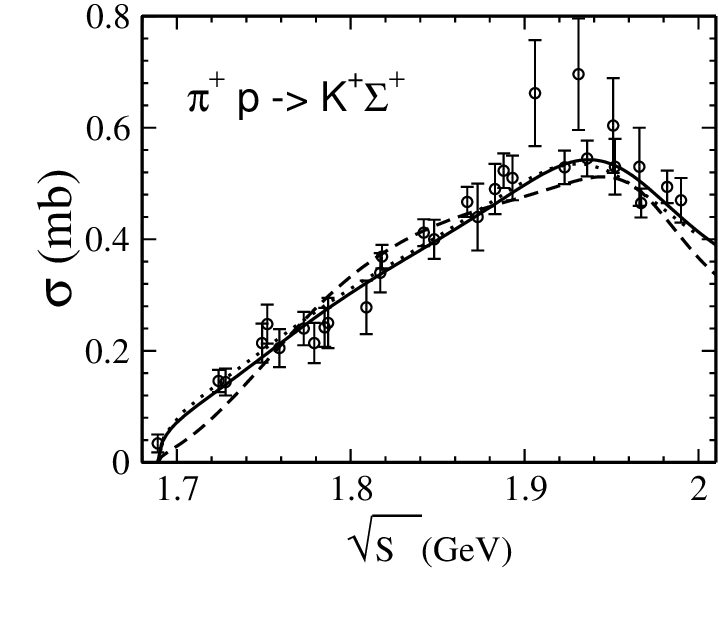}}
       }
       \caption{ 
The total cross sections for the inelastic reactions. 
The solid (dashed) line  corresponds to calculation  C-p-$\pi +(\ffh)$ 
(P-p-$\pi +(\ffh)$) result. The dotted line  shows our previous results C-p-$\pi +$ 
\cite{Penner02}.
For data references see 
\cite{feusti98}.
      \label{crstot}} 
  \end{center}
\end{figure}

\begin{table}[t]
  \begin{center}
    \begin{tabular}
      {l|c|r|r|r|r}
      \hhline{======}
      $L_{2I,2S}$ & mass & $\Gamma_{tot}$ &
      $R_{\pi N}$ & $R_{2\pi N}$ & $R_{K \Sigma}$ \\
      \hhline{======}
      $S_{31}(1620)$ 
& 1614 & 169 &  36.9 & $ 63.1(-)$ & $ 0.91^a$ \\
& 1612 & 175 &  36.0 & $ 64.0(-)$ & $ 0.94^a$ \\
& 1630 & 177 &  42.5 & $ 57.5(+)$ & $ 0.47^a$ \\                     
& 1630 & 177 &  43.4 & $ 56.6(+)$ & $ 0.48^a$ \\
      \hline 

      $S_{31}(1900)^P$ 
& 1987 & 236 &  30.1 & $ 69.8(-)$ & $ 0.1(-)$ \\
& 1984 & 237 &  30.4 & $ 69.5(-)$ & $ 0.1(-)$ \\
      \hhline{======}
      $P_{31}(1750)$ 
& 1752 & 630 &   1.9 & $ 97.4(+)$ & $ 0.7(+)$  \\
& 1752 & 632 &   2.3 & $ 97.2(+)$ & $ 0.6(+)$ \\
& 1978 & 664 &  19.8 & $ 78.9(+)$ & $ 1.3(-)$ \\                    
& 1975 & 676 &  19.5 & $ 79.4(+)$ & $ 1.1(-)$ \\
      \hhline{======}
      $P_{33}(1232)$ 
& 1230 & 104 & 100.0 & $0.001(+)$ & --- \\                 
& 1231 & 101 & 100.0 & $0.002^b(+)$ & --- \\
& 1230 &  94 & 100.0 & $  0.0(+)  $ & ---  \\                     
& 1230 &  94 & 100.0 & $0.000^b(+)$ & --- \\
      \hline 
      $P_{33}(1600)$ 
& 1652 & 244 &  14.3 & $ 85.7(+)$ & $ 0.16^a$ \\                   
& 1652 & 273 &  13.7 & $ 86.3(+)$ & $ 0.22^a$ \\
& 1660 & 371 &  13.7 & $ 86.3(+)$ & $ 0.30^a$ \\                     
& 1656 & 350 &  13.2 & $ 86.8(+)$ & $ 0.28^a$ \\
      \hline 
      $P_{33}(1920)$ 
& 2057 & 514 &  12.5 & $ 82.7(-)$ & $ 4.7(-)$ \\                    
& 2057 & 527 &  15.5 & $ 79.5(-)$ & $ 5.0(-)$ \\
& 2060 & 437 &   8.1 & $ 86.5(-)$ & $ 5.4(-)$ \\                     
& 2056 & 435 &   9.1 & $ 86.8(-)$ & $ 4.1(-)$ \\
      \hhline{======}
      $D_{33}(1700)$ 
& 1677 & 652 &  13.8 & $ 86.2(+)$ & $ 2.11^a$ \\                
& 1680 & 591 &  13.6 & $ 86.4(+)$ & $ 2.09^a$ \\
& 1673 & 671 &  14.6 & $ 85.4(+)$ & $ 3.70^a$ \\                     
& 1674 & 678 &  14.6 & $ 85.4(+)$ & $ 3.68^a$ \\
      \hhline{======}
    \end{tabular}
  \end{center}
  \caption{Properties of  $I=\fth$ resonances considered in the
    present calculation. Masses and total widths $\Gamma_{tot}$ are given 
    in MeV, the decay ratios $R$ in percent of the total width. In
    brackets, the sign of the coupling is given (all $\pi N$ couplings
    are chosen to be positive). $^P$: Only found in the Pascalutsa calculations.
     $^a$: The coupling is given since the resonance is
    below threshold. $^b$: Decay ratio in
    0.1\permil. 1st line: C-p-$\pi +$ ($\ffh$), 
    2nd line: C-p-$\pi +$, 
    3rd  P-p-$\pi +$ ($\ffh$),
    4th line: P-p-$\pi +$. 
    \label{tab32}}
\end{table}

The extracted parameters of the spin-$\ffh$ resonances are presented in Table \ref{52param}. We
note that the total resonance widths  calculated here do not
necessarily  coincide
with the full widths at half maximum because of the energy dependence of
the decay widths (\ref{width}, \ref{helicW}) 
and the  formfactors used \cite{Penner02}. 
 We do not show here the parameters of the $D_{35}(1930)$ resonance because of problems
in the  $D_{35}(1930)$ partial wave (see the previous Section).
 Although a good description 
of the experimental data  is achieved   some differences in  the extracted resonance parameters
for the conventional and the Pascalutsa coupling calculations exist.

We obtain a little lower mass for the  $D_{15}(1675)$ as compared to that  obtained
by Manley and Saleski \cite{manley92} and 
\mbox{Vrana et al. \cite{vrana}}, 
but in agreement with other findings \cite{batinic,cutkosky80}. 
The total width is found 
to be consistent with the results from \cite{cutkosky80,hohler,vrana}. 
In the $\eta N$ channel our calculations show a small ($\approx$0.6\%) decay fraction 
which is somewhat higher than the value obtained by 
Batini\'c et al. : 0.1$\pm 0.1$ \%  \cite{batinic}, whereas Vrana et al. give another
bound: $\pm 1$\%. We conclude that both fits give approximately the same 
results for the resonance masses and branching ratios.

The properties of the $F_{15}(1680)$ state are found to be in good agreement with the 
values recommended by \cite{pdg}. We find a somewhat smaller branching ratio in the $\eta N$
channel as compared to that of \cite{batinic}. However, the obtained value  $R_{\eta N}=$0.1\%
is again  in  agreement with the findings of Vrana et al.\,\cite{vrana}: $\pm 1$\%.
The parameters of the second $F_{15}(2000)$  resonance 
differ strongly in various analyses:
Manley and Saleski give   $490\pm 310$ MeV for the total decay width
while other studies \cite{arndt95,hohler} find 
it at the level of $95-170$ MeV. Moreover, this state has not been identified in the 
investigations of \cite{vrana,batinic}.
Although we find different results for  $\Gamma_{tot}$ in the two independent calculations, 
the branching ratios are close to each other. A small decay 
width of about  4.3\% is found for the $\eta N$ channel (C-p-$\pi +$ ($\ffh$)).  
However, since the $F_{15}(2000)$ resonance is found to be strongly inelastic with  
84-88\% of inelasticity absorbed by the $2\pi N$ channel, the  more 
rich $2\pi N$ data above 1.8 GeV (cf. Fig. \ref{fig2}) are needed for a reliable determination 
of the properties of this state.

The parameters of the $F_{35}(1905)$ state are  in good agreement 
with \cite{pdg}. Both
fits give approximately the same result for the decay branching  ratios.  

Although the extracted resonance masses and total widths can differ for 
the conventional
and the Pascalutsa  descriptions  we find that the branching ratios are almost
identical in both cases. 
Apart from  the $\pi N$ and $2\pi N$ final states we find no 
other significant contributions to the $D_{15}$, $F_{15}$, and $F_{35}$ waves. 
All considered resonances have a  negligible decay ratio to the
$\eta N$, $K\Lambda$, $K \Sigma$, and 
$\omega N$ channels. The only exception is the  $F_{15}(2000)$ resonance
where a small decay  width to $\eta N$  has been found for the conventional coupling 
calculations. Note that there are also small contributions to the $\omega N$ channel. We
corroborate the results of the previous study \cite{Penner02} where the main contribution to this
channel is found to come from the  $P_{11}$ and $P_{13}$ partial waves.

\subsection{The lower spin states}
\label{is12}

In Fig. \ref{crstot} the results for the
$\eta N$, $K \Lambda$, $K \Sigma$, and $\omega N$ total
cross sections are shown. We find a good description of all available experimental data 
including the angle-differential observables 
in these final states. 
The coupled channel effects show up  in a kink structure of
the total $\eta N$,  $K^0\Sigma^0$,  and $K^+\Sigma^-$ cross sections at 1.72 GeV reflecting 
the opening of the $\omega N$ channel. Both calculations give the same results 
for the total cross sections in  all channels except the $\pi^- p \ra\eta n$ reaction 
\cite{Penner02}. 
For comparison our previous best hadronic fit  result C-p-$\pi +$ from \cite{Penner02} 
is shown by 
the dotted line. The calculated cross sections almost coincide with our previous 
results in the conventional coupling calculations. A  visible difference between our 
previous and the new results is seen in the $\omega N$ total cross sections. 
Indeed, only a few datapoints are available above 1.8 GeV so that the $\omega N$ total
cross section is  not strongly constrained by the experimental data. 
This in turn leads to the different $\omega N$ resonance decay widths for the 
$D_{13}(1950)$ state as compared to our previous calculations
\cite{Penner02}.

\begin{table}[t]
  \begin{center}
    \begin{tabular}
      {l|c|r|r|r|r|r|r|r}
      \hhline{=========}
      $L_{2I,2S}$ & mass & $\Gamma_{tot}$ &
      $R_{\pi N}$ & $R_{2\pi N}$ & $R_{\eta N}$ &
      $R_{K \Lambda}$ & $R_{K \Sigma}$ & $R_{\omega N}$ \\
      \hhline{=========}
      $S_{11}(1535)$ 
& 1540 & 156 & 35.7 & $11.2(+)$ & $53.1(+)$ & $ 0.02^a$ & $-2.54^a$ & --- \\
& 1542 & 148 & 37.7 & $11.5(+)$ & $50.8(+)$ & $ 0.02^a$ & $ 0.27^a$ & --- \\
& 1548 & 125 & 34.4 & $ 0.4(-)$ & $65.2(+)$ & $-4.46^a$ & $ 0.43^a$ & --- \\
& 1545 & 117 & 36.6 & $ 0.9(-)$ & $62.6(+)$ & $-4.46^a$ & $ 0.26^a$ & --- \\
      \hline
      $S_{11}(1650)$ 
& 1676 & 161 & 65.4 & $19.6(+)$ & $ 6.4(-)$ & $ 8.6(-)$ & $-0.53^a$ & --- \\
& 1671 & 158 & 65.1 & $22.7(+)$ & $ 5.1(-)$ & $ 7.1(-)$ & $-0.54^a$ & --- \\
& 1703 & 294 & 68.0 & $14.3(-)$ & $ 4.9(+)$ & $12.8(-)$ & $-0.36^a$ & --- \\
& 1699 & 276 & 68.2 & $14.7(-)$ & $ 3.8(+)$ & $13.3(-)$ & $-0.50^a$ & --- \\
      \hhline{=========}
      $P_{11}(1440)$ 
& 1508 & 571 & 60.7 & $39.3(+)$ & $ 3.51^a$ & $ 3.43^a$ & $-2.22^a$ & --- \\
& 1490 & 463 & 61.5 & $38.5(+)$ & $ 3.27^a$ & $ 3.43^a$ & $-1.01^a$ & --- \\
& 1516 & 644 & 60.5 & $39.5(+)$ & $ 3.17^a$ & $ 1.97^a$ & $ 3.67^a$ & --- \\
& 1515 & 639 & 60.6 & $39.4(+)$ & $ 4.17^a$ & $ 1.97^a$ & $ 3.64^a$ & --- \\
      \hline
      $P_{11}(1710)$ 
& 1753 & 534 &  2.3 & $30.7(+)$ & $26.2(-)$ & $ 0.1(+)$ & $20.8(-)$ & 19.9 \\
& 1770 & 430 &  2.0 & $42.7(+)$ & $31.6(-)$ & $ 0.9(+)$ & $ 6.3(-)$ & 16.4 \\
& 1704 & 380 &  7.8 & $27.8(-)$ & $35.6(+)$ & $26.6(-)$ & $ 2.3(-)$ & ---  \\                     
& 1701 & 348 &  8.5 & $25.7(-)$ & $38.3(+)$ & $26.3(-)$ & $ 1.3(-)$ & --- \\
      \hhline{=========}
      $P_{13}(1720)$ 
& 1725 & 267 & 12.2 & $66.2(+)$ & $ 2.1(-)$ & $ 8.1(-)$ & $10.6(-)$ &  0.8 \\
& 1724 & 295 & 15.4 & $65.2(+)$ & $ 1.2(+)$ & $ 9.9(-)$ & $ 7.5(-)$ &  0.7 \\
& 1694 & 170 & 15.8 & $82.7(+)$ & $ 0.1(+)$ & $ 1.1(+)$ & $ 0.4(+)$ & ---  \\             
& 1700 & 148 & 14.2 & $83.1(+)$ & $ 0.0(+)$ & $ 1.7(+)$ & $ 1.0(+)$ & --- \\
      \hline
      $P_{13}(1900)$ 
& 1962 & 700 & 24.7 & $52.8(-)$ & $ 9.3(+)$ & $ 3.5(-)$ & $ 0.1(+)$ &  9.6 \\
& 1962 & 683 & 19.1 & $58.2(-)$ & $11.9(+)$ & $ 1.9(-)$ & $ 0.8(+)$ &  8.1 \\
& 1948 & 792 & 23.5 & $51.7(-)$ & $ 8.3(+)$ & $ 0.8(+)$ & $ 0.3(+)$ & 15.5 \\                 
& 1963 & 694 & 15.7 & $58.2(-)$ & $ 3.0(+)$ & $ 0.1(+)$ & $ 0.0(+)$ & 22.9 \\
      \hhline{=========}
      $D_{13}(1520)$ 
& 1508 &  91 & 58.9 & $41.1(-)$ & $ 1.4^b(+)$ & $ 0.44^a$ & $ 1.47^a$ & --- \\
& 1512 &  95 & 58.7 & $41.3(-)$ & $ 3.1^b(+)$ & $ 0.44^a$ & $ 1.20^a$ & --- \\
& 1509 &  93 & 60.7 & $39.3(-)$ & $ 1.5^b(+)$ & $ 0.86^a$ & $-3.32^a$ & --- \\     
& 1509 &  91 & 60.1 & $39.9(-)$ & $ 2.2^b(+)$ & $ 0.86^a$ & $-3.23^a$ & --- \\
      \hline

      $D_{13}(1700)^P$ 
& 1743 &  67 &  0.8 & $31.5(+)$ & $ 4.4(+)$ & $ 4.3(-)$ & $ 1.4(-)$ & 57.6 \\
& 1745 &  55 &  1.6 & $43.4(+)$ & $ 1.7(+)$ & $ 6.7(-)$ & $ 1.2(-)$ & 45.3 \\
      \hline
      $D_{13}(1950)$ 
& 1927 & 855 & 15.5 & $33.2(+)$ & $ 0.4(-)$ & $ 1.2(+)$ & $ 2.6(+)$ & 47.0 \\
& 1946 & 885 & 16.2 & $49.1(+)$ & $ 2.2(-)$ & $ 1.2(+)$ & $ 1.9(+)$ & 29.4 \\
& 1946 & 703 & 14.1 & $56.0(+)$ & $ 0.0(+)$ & $ 2.0(-)$ & $ 0.8(+)$ & 27.1 \\                     
& 1943 & 573 & 13.3 & $50.8(+)$ & $ 0.0(-)$ & $ 2.2(-)$ & $ 0.7(+)$ & 32.9 \\
      \hhline{=========}
    \end{tabular}
  \end{center}
  \caption{Properties of $I=\foh$ resonances considered in the
    calculation. Notation is the same as in Table \ref{tab32}.
    \label{tab12}} 
\end{table}
%%%%%%%%%%%%%%%%%%%%%%%%%
\begin{table}[t]
  \begin{center}
    \begin{tabular}
      {l|l|l|l|l|r}
      \hhline{======}
      $L_{2I,2S}$ & mass & $\Gamma_{tot}$ &
      $R_{\pi N}$ & $R_{2\pi N}$ & $R_{K \Sigma}$ \\
      \hhline{======}
      $S_{31}(1620)$ & 1614    & 169 &  36.9    & 63   &  \\
                     & 1620    & 150 & 25(5)    & 75(5)& \\
                     & 1672(7) & 154(37 )& 9(2) &      & \\
                     & 1617(15)& 143(42) & 45(5)&      & \\
      \hline 
      $S_{31}(1900)^P$ & 1987 & 236 &  30 & 70 & 0.1  \\
                     & 1900 & 200 & 20(10) & & \\
                     & 1920(24) & 263(39)& & 41(4) & \\
                     & 1802(87) & 48(45) & 33(10) & \\
      \hhline{======}
      $P_{31}(1750)$ & 1752    & 630     & 1.9  &  97.4 & 0.7  \\
                     & 1750    & 300     & 8(3) &       & \\
                     & 1744(36)& 300(120)& 8(3) &       & \\
                     & 1721(61)& 70(50)  & 6(9) &       & \\
      \hhline{======}
      $P_{33}(1232)$ & 1230 & 104 & 100.0 & 0.001  &  \\   
                     & 1232 & 120 & $>99$ & 0 & \\
                     & 1231(1) & 118(4) & $\;\;$100 & & \\
                     & 1234(5) & 112(18) & $\;\;$100(1) & \\
      \hline 
      $P_{33}(1600)$ & 1652     & 244     &  14.3  & 86 & $  $ \\ 
                     & 1600     & 350     &  18(7) & 82(8) & \\
                     & 1706(10) & 430(73) &  12(2) &       & \\
                     & 1687(44) & 493(75) &  28(5) &       & \\
      \hline 
      $P_{33}(1920)$ & 2057     & 514    & 12.5      & 82.7 &  4.7 \\
                     & 1920     & 200    & 13(7)     &      & \\
                     & 2014(16) & 152(55)& $\;\,$2(2)&      & \\
                     & 1889(100)& 123(53)& $\;\,$5(4)&      &\\
      \hhline{======}
      $D_{33}(1700)$ & 1677    & 652     &  14       & 86   & \\   
                     & 1700    & 300     & 15(5)     & 85(5)& \\
                     & 1762(44)& 600(250)&14(6)      &      & \\
                     & 1732(23)& 119(70) & $\;\,$5(1)& \\
      \hhline{======}
      $F_{35}(1905)$ & 1859    & 400     &  11       & 89   & $0.7^b$\\   
                     & 1905    & 350     & 10(5)     & 90(5)& \\
                     & 1881(18)& 327(51) & 12(3)     &      & \\
                     & 1873(77)& 461(111)& $\;\,$9(1)& \\

      \hhline{======}
    \end{tabular}
  \end{center}
  \caption{Properties of $I=\fth$ resonances for
    the calculation C-p-$\pi +$ ($\ffh$) (1st line) in comparison with the
    values from \cite{pdg} (2nd line), \cite{manley92} (3rd line), and 
    \cite{vrana} (4th line). In brackets, the estimated
    errors are given. The mass and total width are given in MeV, the
    decay ratios in percent.
    $^b$: The decay ratio is given in 0.1\permil. $^P$:
    Calculation P-p-$\pi +$ ($\ffh$). \label{comp32}} 
\end{table}
\begin{table}[t]
  \begin{center}
    \begin{tabular}
      {l|l|l|l|l|l|l|c|c}
      \hhline{=========}
      $L_{2I,2S}$ & mass & $\Gamma_{tot}$ &
      $R_{\pi N}$ & $R_{2\pi N}$ & $R_{\eta N}$ &
      $R_{K \Lambda}$ & $R_{K \Sigma}$ & $R_{\omega N}$ \\
      \hhline{=========}
      $S_{11}(1535)$ & 1540 & 156 & 36 & 11 & 53 &   &  & \\
                     & 1535 & 150 & 45(10) &  $\;\,$6(5) & 43(12) & & & \\
                     & 1534(7) & 151(27) & 51(5) & & & & & \\
                     & 1542(3) & 112(19) & 35(8) & & 51(5)& & & \\
      \hline 
      $S_{11}(1650)$ & 1676    & 161    & 65    & 20   & 6   &   9 &  &  \\
                     & 1650    & 150    & 72(17)& 15(5)& 6(3)& 7(4)&  & \\
                     & 1659(9) & 173(12)& 89(7) &      &     &     &  & \\
                     & 1689(12)& 202(40)& 74(2) &      & 6(1)&     &  & \\
      \hhline{=========}
      $P_{11}(1440)$ & 1508 & 571 & 61 & 39 &  &  &  &  \\
                     & 1440 & 350 & 65(5) & 35(5) & & & & \\
                     & 1462(10) & 391(34) & 69(3) &  & & & & \\
                     & 1479(80) & 490(120) & 72(5) & &0(1) & & & \\
      \hline 
      $P_{11}(1710)$ & 1753     & 534      &  2    & 31    & 26 &  0.1 & 21 & 20 \\
                     & 1710     & 100      & 15(5) & 65(25)&    & 15(10) & & \\
                     & 1717(28) & 480(230) & 9(4)  &       &    &        & & \\
                     & 1699(65) & 143(100) & 27(13) &       & 6(1)&       & & \\
      \hhline{=========}
      $P_{13}(1720)$ & 1725 & 267 & 12 & 66 &  2 &  8 & 11 &  1 \\
                     & 1720      & 150      & 15(5)      & $>70$ & 4 & $\;\,$8(7) & & \\
                     & 1717 (31) & 380(180) & 13(5)      & & 4 & $\;\,$1 & & \\
                     & 1716(112) & 121(39)  & $\;\,$5(5) & & 4(1) & & & \\
      \hline 
      $P_{13}(1900)$ & 1962 & 700 & 25 & 52 &  9 &  4 & 0.1 & 10 \\
                     & 1900 & 500 &  &  & & & \\
                     & 1879(17) & 498(78) & 26(6)& & & & & \\
                     & NF & & & & & & & \\
      \hhline{=========}
      $D_{13}(1520)$ & 1508    &  91 & 59 & 41 & $ 1^b$ &  &  & \\
                     & 1520    & 120    & 55(5) & 45(5)    &    &  & & \\
                     & 1524(4) & 124(8) & 59(3) &         &    &  & &\\
                     & 1518(3) & 124(4) & 63(2) &     & 0(1)   &  & &\\
      \hline 
      $D_{13}(1700)^P$ & 1743 &  67 &  1 & 32 &  4 &  4 &  1 & 58 \\
                       & 1700 & 100 & 10(5) & 90(5) & & $<3$ & & \\
                       & 1737(44) & 250(220) & 1(2) & & & & & \\
                       & 1736(33) & 175(133) & $\;\,$4(2) & & 0(1) & & & \\
      \hline 
      $D_{13}(1950)$ & 1927 & 855 & 16 & 33 &  0.4 &  1 &  3 & 47 \\
                     & 2080     &           &       &    &        &  & & \\
                     & 1804(55) &  450(185) & 23(3) &    &        &  & & \\
                     & 2003(18) & 1070(858) & 13(3) &    & 0(2)   &  & & \\
      \hhline{=========}
      $D_{15}(1675)$ & 1665     & 144       & 40    & 59   &  1   & 0    &  & \\
                     & 1675     & 150       & 45(5) & 55(5)& 0(1) &$<1$ &  & \\
                     & 1676(2)  & 159(7)    & 47(2) &      &      &     &  & \\
                     & 1685(4)  & 131(10)   & 35(1) &      & 0(1) &     &  & \\
      \hhline{=========}
      $F_{15}(1680)$ & 1674    & 120    & 69    & 32   &  0.1 &  0 &   & \\
                     & 1680    & 130(10)& 65(5) & 35(5)&  0(1)&    &   & \\
                     & 1684(4) & 139(8) & 70(3) &      &      &    &   & \\
                     & 1679(3) & 128(9) & 69(2) &      &  0(1)&    &   & \\
      \hline
      $F_{15}(2000)$ & 1981     & 361      &  9 & 84 & 4 & $  6.8^b $ &  0.4 &  2 \\
                     & 2000     &          &    &    &   &               &      &   \\
                     & 1903(87) & 490(310) &  8(5) &     &   &   &      &   \\
                     & NF       &          &       &     &   &   &      &   \\
      \hhline{=========}
    \end{tabular}
  \end{center}
  \caption{Comparison of $I=\foh$ resonance properties. Notation as in
    Table \ref{comp32}.
    \label{comp12}}
\end{table}

The extracted lower spin resonance properties are hardly changed in 
comparison with the previous findings \cite{Penner02}.
However, the inclusion of the
spin-$\ffh$ states leads to some modification of the isospin-$\foh$ baryon resonance
parameters. 
We discuss these states only when apparent differences with earlier 
studies are found. 

The results for isospin-$\fth$($\foh$) resonance parameters are presented in 
Table \ref{tab32}\,\refe{tab12}. In Tables\,\ref{comp32} and \ref{comp12} a comparison 
with the values given by the partical data group \cite{pdg} and parameters
from \cite{manley92,vrana} are shown.
All resonance states investigated in our previous work \cite{Penner02} were included in the present 
calculations. We corroborate all the  resonances identified  in \cite{Penner02}.    
One sees that the inclusion of spin-$\ffh$ resonances hardly affects the results for the 
isospin-$\fth$ resonance properties.  

Some deviations in comparison with our previous studies
are found for the isospin-$\foh$ resonances, see Table \ref{tab12}. 
We obtain a somewhat larger width for the Roper resonance
$P_{11}(1440)$ in the new C-p-$\pi+$($\ffh$) calculations. Indeed the properties 
of this state are found to 
be very sensitive to background contributions \cite{Penner02,pm2}. The inclusion 
of the spin-$\ffh$ states gives an additional lower-spin background  which 
leads to the increase of the total width for the conventional spin-$\ffh$ coupling
calculation
in comparison with that  obtained in the previous study \cite{Penner02}. 
The branching ratios are found to be in agreement with the values recommended
in  \cite{pdg}. 

The  $P_{11}(1710)$ resonance 
has a three-star status  and its  properties are not completely established 
\cite{pdg}. The parameters of this state are also found to be strongly influenced by 
the lower spin off-shell contributions from the spin-$\ffh$ resonances in the conventional
coupling calculations. 
While the $\pi N$ decay width  remains almost unchanged, the additional background affects the 
inelactic channels and the fit moves the strength from $2\pi N$ and $\eta N$ to the $K\Sigma$ and 
$\omega N$ final states. This also leads to the decrease of the  $\Lambda_\foh$ 
(see Table \ref{tabcutoff}).
 The resulting $K\Sigma$ decay width turns out to be rather large: 
$R_{K\Sigma}$=21\%. 
We conclude that the
properties of the $P_{11}(1710)$ state  are rather sensitive to
background contributions in the conventional coupling calculations. 
While the conventional couplings lead to  different resonance properties in the
$P_{11}$ partial wave as compared to our previous study the resonance parameters 
in the Pascalutsa coupling calculations are only slightly changed. The $P_{11}(1710)$
resonance width $\Gamma_{tot}$=534(430) MeV  obtained in the conventional (Pascalutsa) coupling 
calculation is in  agreement with the  findings  \cite{manley92} 
(see Table \ref{comp12}) but smaller than  the results from 
\cite{vrana,arndt96,cutkosky90}. We also find the $\eta N$ ratio about 25-30\%, close to  the 
result of  Batini\'c et al. \cite{batinic}.  

There are two resonances  (\,$ P_{13}(1720)$ and $ P_{13}(1900)$\,) 
included in the $P_{13}$ wave. 
It is interesting to look at the results for the $ P_{13}(1900)$ resonance in the
Pascalutsa coupling calculations. Although the Pascalutsa spin-$\ffh$ coupling does not have
any lower-spin off-shell background and therefore does not directly affect the $\pi N$ and
$2\pi N$ lower-spin partial waves, the inclusion of the spin-$\ffh$ resonances gives  
additional contributions to the $\eta N$, $K\Lambda$,$K\Sigma$, and $\omega N$ channels.
The fit has tried to compensate these changings  giving also  somewhat different values for the  
$\pi N$ and $2\pi N$ decay widths as compared to our previous results. 
The same effect is seen in the properties of the $D_{13}(1700)$ resonance where
noticeable changes in the  $2\pi N$ and $\omega N$ channels are also observed 
(cf. Table \ref{tab12}). Note, that although  the resonance mass of the $P_{13}(1900)$ is rather 
well  fixed here, the inclusion of photoproduction data may change the situation  
\cite{Penner02,pm2}. 

As in the case of the $P_{11}(1710)$ resonance, the properties of the   $D_{13}(1950)$
state are also influenced by the spin-$\ffh$ off-shell contributions in the conventional 
coupling calculations. The most striking difference between our previous and the new results
for the $D_{13}(1950)$ state  is  found in the $2\pi N$ and $\omega N$ channels. Here 
the $2\pi N$-flux is moved to the $\omega N$ channel. On the other hand, the investigations of 
photo-induced  reactions \cite{pm2} show that the $D_{13}$ resonances play a 
minor role in $\omega N$  when the photoproduction data are included. We corroborate our 
previous findings that  using only $\pi N \to \omega N$ is  not sufficient for a reliable
determination of the  $\omega N$ strength in each partial wave.  

Finally, we conclude that the inclusion of the spin-$\ffh$ resonances leads to
some modifications of the resonance parameters extracted in our previous study \cite{Penner02}. 
While the properties of the isospin-$\fth$ states  are hardly changed, the parameters of 
the isospin-$\foh$ resonances differ in some specific cases from those  extracted 
in \cite{Penner02}. The largest changes of the decay ratios can be observed  in the $2\pi N$, 
$\eta N$, $K\Sigma$, and 
$\omega N$ channels, see Table \ref{tab12}. As stressed in \cite{Penner02} 
photoproduction data are  inevitable for a reliable fixing of the resonance parameters.

The cutoff values obtained in the different calculations are shown in Table \ref{tabcutoff}.
Except for $\Lambda_\foh$ in the conventional coupling calculations, the results 
for the cutoff parameters remain unchanged in comparison with the previous findings.
The decrease in this cutoff is caused by the influence
of the additional spin-$\foh$ background contributions from the newly incorporated 
spin-$\ffh$ resonances. Indeed, as  discussed above, the inclusion of 
these states gives strong contributions to the $P_{11}$ partial wave which 
increases the $P_{11}(1440)$ and the  $P_{11}(1710)$ total widths. 
This effect leads to the modification of the spin-$\foh$ cuff-off giving a somewhat lower
value as compared to our previous results.
The large difference in  $\Lambda_t$  between the conventional and the Pascalutsa coupling
calculations is related to the need for a larger  $t$-channel
non-resonant background in case of the Pascalutsa coupling calculations (see discussion 
in \cite{Penner02}). 
\begin{table}[t]
  \begin{center}
    \begin{tabular}
      {l|r|r|r|r|r}
      \hhline{======}&
      $\Lambda_N$ [GeV] & 
      $\Lambda_\foh$ [GeV] & 
      $\Lambda_\fth$ [GeV] & 
      $\Lambda_\ffh$ [GeV] & 
      $\Lambda_t$ [GeV] \\
      \hhline{======}
 C-p-$\pi+$($\ffh$)& 1.16 & 2.79 & 1.03 & 1.16 & 0.71 \\
 C-p-$\pi+$     & 1.16 & 3.64 & 1.04 & ---  & 0.70 \\
 P-p-$\pi+$($\ffh$)& 1.17 & 4.30 & 1.02 & 0.95 & 1.82 \\
 P-p-$\pi+$     & 1.17 & 4.30 & 1.02 & ---  & 1.80 \\
      \hhline{======}
    \end{tabular}
  \end{center}
  \caption{Results for the formfactor cutoff values for the formfactors in 
    comparison with the previous findings.
    The lower index denotes
    the intermediate particle, i.e. $N$: nucleon, $\foh$:
    spin-$\foh$ resonance, $\fth$: spin-$\fth$, $\ffh$: spin-$\ffh$ resonance, $t$:
    $t$-channel meson.
    \label{tabcutoff}}
\end{table}

\section{Summary and outlook}

We have performed a first investigation of the pion-induced reactions on the nucleon
within the effective Lagrangian
$K$-matrix  approach which includes the  higher spin-$\ffh$ resonances.
To investigate the  influence of additional background from the 
spin-$\fth$ and -$\ffh$ resonances 
calculations using the conventional and Pascalutsa higher-spin couplings have been carried
out. A good description of the available  experimental  data has been achieved in all 
$\pi N$, $2\pi N$, $\eta N$, $K \Lambda$, $K\Sigma$, and $\omega N$ final states within both  
frameworks giving however a somewhat worse  $\chi^2$ for the Pascalutsa coupling prescription. 
Apart from $2\pi N$ we find no significant contributions  from  other channels to the total 
$\pi N$ inelasticities in the spin-$\ffh$ waves.
Nevertheless, the inclusion of the higher-spin states  improves the 
$\chi^2$ in almost all channels. 

We have also found  evidence for the $F_{15}(2000)$ resonance which is rated two-star by \cite{pdg} 
and has not been included in the most recent resonance analysis by Vrana et al. \cite{vrana}.
The small $\pi N$ decay ratio shows the necessity for more precise $2\pi N$ data 
to identify this state more reliably in the purely hadronic calculations. 
For most resonances the extracted parameters for the lower-spin 
states have only slightly changed in the new calculations 
compared to those obtained in 
our previous  study \cite{Penner02}. 
However, for some resonances  deviations from  \cite{Penner02} 
for the pure hadronic fits are observed. 
This underlines the necessity of the inclusion of  photoproduction data to fix the resonance
couplings with certainty. The extracted cutoff parameters are also in general identical 
to those  obtained in the previous  study.

We are proceeding with the extension of our model by performing a combined analysis of
pion- and photon-induced reactions taking into account spin-$\ffh$ states.
Moreover the decomposition of the $2\pi N$ channel into  intermediate $\rho N$, 
$\pi \Delta$ etc. states will be the subject of further investigations.

\end{document}